\documentclass[conference,letter]{IEEEtran}
\usepackage{amsthm}
\usepackage{amsmath}
\usepackage{amssymb} 
\usepackage{amsmath}
\usepackage{amsfonts}
\usepackage{algorithmic}
\usepackage{graphicx}
\usepackage{cite}

\usepackage{subcaption}
\usepackage{textcomp}
\usepackage{xcolor}
\usepackage{tabularx}
\usepackage{lipsum}
\usepackage{float}
\usepackage{stmaryrd}
\usepackage{flushend}
\usepackage{multirow}

\usepackage{url}
\usepackage{ifdraft}
\usepackage{multirow}
\usepackage{hyperref}

\usepackage{footmisc}

\hypersetup{
    unicode=false,          % non-Latin characters in Acrobat's bookmarks
    pdftoolbar=true,        % show Acrobat's toolbar?
    pdfmenubar=true,        % show Acrobat's menu?
    pdffitwindow=false,     % window fit to page when opened
    pdfstartview={FitH},    % fits the width of the page to the window
    pdftitle={The Title of the Paper},
    pdfauthor={Dr. Ferkan Yilmaz},% author
    pdfsubject={Subject},   % subject of the document
    pdfcreator={Creator},   % creator of the document
    pdfproducer={latex.exe + dvips.exe + ps2pdf.exe}, % producer of the document
    pdfkeywords={XXXX, XXXX, XXXX, XXXX.}, % list of keywords
    pdfnewwindow=true,      % links in new window colorlinks=false, % false: boxed links; true: colored links
    linkcolor=red,          % color of internal links
    citecolor=green,        % color of links to bibliography
    filecolor=magenta,      % color of file links
    urlcolor=cyan           % color of external links
}

\newcommand{\argmax}{\mathop{\mathrm{argmax}}\limits}

\newcommand{\IEEEcompact}{
\fontdimen3\font=0.2ex% inter word stretch
\fontdimen2\font=0.5ex% inter word space
}

\theoremstyle{plain}

\theoremstyle{plain}

\theoremstyle{plain}

\theoremstyle{plain}

\theoremstyle{plain}

\newcommand{\figref}[1]{Fig.\,\protect\ref{#1}}

\newcommand{\secref}[1]{Section\,\protect\ref{#1}}

\newcommand{\ITMsymbol}{$\hfill\square$}

\def\BibTeX{{\rm B\kern-.05em{\sc i\kern-.025em b}\kern-.08em
    T\kern-.1667em\lower.7ex\hbox{E}\kern-.125emX}}
\usepackage{hyphenat}
\IEEEoverridecommandlockouts 
\begin{document}
%-------------------------------------------------------------------------------------------------------------------
\IEEEcompact
%%%%%%%%%%%%%%%%%%%%%%%%%%%%%%%%%%%%%%%%%%%%%%%%%%%%%%%%%%%%%%%%%%%%%%%%%%%%%%%%%%%%%%%%%%%%%%%%%%%%%%%%%%%%%%%%%%%%%%%%%%%%
%% ___________.__  __  .__          
%% \__    ___/|__|/  |_|  |   ____  
%%   |    |   |  \   __\  | _/ __ \ 
%%   |    |   |  ||  | |  |_\  ___/ 
%%   |____|   |__||__| |____/\___  >
%%                               \/ 
% \title{A Novel GRAND-Based Decode-and-Forward Multihop Transmission for 5G and 5G-Beyond Networks\\
% \title{Unlocking 6G Potential: Integrating Multi-Hop, CRC, and GRAND for Enhanced Wireless Communication\\
%\title{Unlocking 6G Potential: Multi-Hop Transmission, CRC, and GRAND Integration in Future Wireless Networks\\
% \title{Beyond Boundaries: Multihop Transmission, CRC, and GRAND Integration in Future Wireless 6G Networks\\ 
% \title{GRAND Design for 6G: Multi-hop and CRC Integration for Future Networks\\
% \title{Beyond Traditional Approaches: Unveiling the Power of GRAND for 6G Networks\\
% \title{Unlocking 6G Potential: Integrating Multi-Hop, CRC, and GRAND for Enhanced Wireless Networks\\
% \title{Beyond 5G/6G: Unveiling the Power of GRAND for Reliable Multihop Transmissions\\
%\title{Beyond 5G/6G: Unveiling the Power of Multihop Transmission, CRC, and GRAND Integration in Future Wireless Networks\\
% \title{Unlocking 5G-Beyond/6G Potential: Integrating Multihop, CRC, and GRAND for Enhanced Wireless Networks\\
\title{Unlocking Potential: Integrating Multihop, CRC, and GRAND for Wireless 5G-Beyond/6G Networks
\thanks{This work is supported by Istanbul Technical University BAPS\.{I}S MAB-2023-44565 project. The publication is funded by T\"{U}B\.{I}TAK-B\.{I}LGEM. \vspace{-6mm}}
}
\author{
\IEEEauthorblockN{
  	${}^{(a)}$Bora Bozkurt,
        ${}^{(b)}$Emirhan Zor, and 
        ${}^{(c)}$Ferkan Yilmaz}\\[-3mm]
  	\IEEEauthorblockA{
        ${}^{(a,b,c)}${Istanbul Technical University, Electronics and Communication Engineering, Istanbul, T\"{u}rkiye}\\
       ${}^{(a)}${Communications and Signal Processing Research (H\.{I}SAR) Laboratory, T\"{U}B\.{I}TAK B\.{I}LGEM, 41470, Kocaeli, T\"{u}rkiye}\\        
            \{\!
                ${}^{(a)}$\texttt{bozkurtb19},
                ${}^{(b)}$\texttt{zor18},
                ${}^{(c)}$\texttt{yilmazf}
            \!\}\texttt{@itu.edu.tr}
  	}
%------------------------------------------------------------
\vspace{-6mm}
%------------------------------------------------------------
}
\IEEEoverridecommandlockouts
\IEEEpubid{\makebox[\columnwidth]{979-8-3503-8481-9/24/\$31.00 ©2024 IEEE \hfill}
\hspace{\columnsep}\makebox[\columnwidth]{ }}
\maketitle
\IEEEpubidadjcol
\IEEEcompact
%%%%%%%%%%%%%%%%%%%%%%%%%%%%%%%%%%%%%%%%%%%%%%%%%%%%%%%%%%%%%%%%%%%%%%%%%%%%%%%%%%%%%%%%%%%%%%%%%%%%%%%%%%%%%%%%%%%%%%%%%%%%
%%   _____ ___.             __                        __
%%  /  _  \\_ |__   _______/  |_____________    _____/  |_
%% /  /_\  \| __ \ /  ___/\   __\_  __ \__  \ _/ ___\   __\
%%/    |    \ \_\ \\___ \  |  |  |  | \// __ \\  \___|  |
%%\____|__  /___  /____  > |__|  |__|  (____  /\___  >__|
%%        \/    \/     \/                   \/     \/
\begin{abstract}
As future wireless networks move towards millimeter wave (mmWave) and terahertz (THz) frequencies for 6G, multihop transmission using Integrated Access Backhaul (IABs) and Network-Controlled Repeaters (NCRs) will be highly essential to overcome coverage limitations. This paper examines the use of Guessing Random Additive Noise (GRAND) decoding for multihop transmissions in 3GPP networks. We explore two scenarios: one where only the destination uses GRAND decoding, and another where both relays and the destination leverage it. Interestingly, in the latter scenario, the Bit Error Rate (BER) curves for all hop counts intersect at a specific Signal-to-Noise Ratio (SNR), which we term the GRAND barrier. This finding offers valuable insights for future research and 3GPP standard development. Simulations confirm the effectiveness of GRAND in improving communication speed and quality, contributing to the robustness and interconnectivity of future wireless systems, particularly relevant for the migration towards mmWave and THz bands in 6G networks. Finally, we investigate the integration of multihop transmission, CRC detection, and GRAND decoding within 3GPP networks, demonstrating their potential to overcome coverage limitations and enhance overall network performance.
\end{abstract}

% Alternative abstract:
% As future wireless networks move towards millimeter wave (mmWave) and terahertz (THz) frequencies for 6G, multihop transmission using Integrated Access Backhaul (IABs) and Network-Controlled Repeaters (NCRs) are expected to be highly essential to overcome coverage limitations. This paper examines the use of Guessing Random Additive Noise (GRAND) decoding for multihop transmissions in 3GPP networks. We explore and provide BER-cause analysis for two scenarios: one where only the destination uses GRAND decoding, and another where both the relay nodes and the destination leverage it. For both scenarios, simulations confirm the effectiveness of GRAND in improving communication quality, contributing to the robustness and inter-connectivity of future wireless systems, which is particularly relevant for the migration towards mmWave and THz bands in 6G networks. Interestingly, in the latter scenario, the Bit Error Rate (BER) curves for all hop counts intersect at a specific Signal-to-Noise Ratio (SNR), which we term the GRAND barrier. This finding offers valuable insights for future research and 3GPP standard development.  

%%  ____  __.                                    .___      
%% |    |/ _|____ ___.__.__  _  _____________  __| _/______
%% |      <_/ __ <   |  |\ \/ \/ /  _ \_  __ \/ __ |/  ___/
%% |    |  \  ___/\___  | \     (  <_> )  | \/ /_/ |\___ \ 
%% |____|__ \___  > ____|  \/\_/ \____/|__|  \____ /____  >
%%         \/   \/\/                              \/    \/ 
% Note that keywords are not normally used for peerreview papers.

%%%%%%%%%%%%%%%%%%%%%%%%%%%%%%%%%%%%%%%%%%%%%%%%%%%%%%%%%%%%%%%%%%%%%%%%%%%%%%%%%%%%%%%%%%%%%%%%%%%%%%%%%%%%%%%%%%%%%%%%%%%%
%% .___        __                    .___             __  .__               
%% |   | _____/  |________  ____   __| _/_ __   _____/  |_|__| ____   ____  
%% |   |/    \   __\_  __ \/  _ \ / __ |  |  \_/ ___\   __\  |/  _ \ /    \ 
%% |   |   |  \  |  |  | \(  <_> ) /_/ |  |  /\  \___|  | |  (  <_> )   |  \
%% |___|___|  /__|  |__|   \____/\____ |____/  \___  >__| |__|\____/|___|  /
%%          \/                        \/           \/                    \/ 
\section{Introduction} \label{Sec:Introduction}
\IEEEcompact
As network frequencies progress towards mmWave and THz bands for 6G, addressing coverage challenges necessitates the deployment of various network nodes such as Integrated Access Backhaul (IABs) \cite{3GPPTR38874} and Network-Controlled Repeaters (NCRs) \cite{3GPPTR38867}. Multihop transmission, a technique already utilized in LTE-Advanced and 5G, is expected to become increasingly crucial in 6G networks \cite{3GPPTS38401}. This technique, outlined in \cite{3GPPTS38401}, not only optimizes network performance but also enhances connectivity \cite{akyildiz2022terahertz,shen2023five,bhat20216g}.
The incorporation of multihop transmission in 6G networks will further enhance network performance and connectivity \cite{tataria20216g}. By employing relay nodes, multihop extends coverage to remote areas, enhances capacity in dense environments, and improves reliability and latency by providing alternative transmission paths. In 5G, the concept of IAB \cite{3GPPTS38401,3GPPTR38874} facilitates flexible multihop deployments, particularly crucial for applications like V2X communication. The IAB node, introduced in NR Release 16, acts as a decode-and-forward relay with functionality up to Layer-2 \cite{madapatha2020integrated,zhai2023interference,flamini2022toward}. The integration of multihop transmission in IAB deployments has already begun to improve network efficiency and address coverage challenges in both urban and rural areas \cite{akyildiz2022terahertz,tezergil2022wireless}. Another network node introduced in Release-18, the NCR, represents an evolution beyond conventional amplify-and-forward RF repeaters. Unlike traditional repeaters, the NCR incorporates intelligence to decode control information, enhancing its functionality without interfering with user equipment (UE) communications \cite{3GPPTR38867}.
With ongoing technological advancements and evolving network architectures, 6G stands to fully exploit the advantages of multihop transmission, contributing to a more resilient and interconnected wireless ecosystem. The seamless integration of relay nodes and multihop communication will be pivotal in unlocking the full potential of 6G networks, revolutionizing future connectivity and communication experiences \cite{akyildiz2022terahertz,shen2023five,tataria20216g}. As 6G endeavors to expand connectivity, improve network efficiency, and support innovative applications, widely common conjecture noticeably anticipates that multihop transmission will play a prominent role in achieving these objectives.

%%
%% New Paragraph
%%
It is also worth noticing that Cyclic Redundancy Check (CRC) is crucial for detecting errors in IABs\cite{3GPPTR38874} and NCRs \cite{3GPPTR38867} within 3GPP networks, where detailed CRC specifications, including the generator polynomial and length, are provided in 3GPP technical documentation\cite{3GPPTR38212}. As such, CRC lengths can be configured as 24, 16, 12, 8, or 0 bits not only to identify transmission errors caused by noise, interference, or channel impairments  but also to facilitate the retransmission of faulty packets and improving overall network reliability and Quality of Service (QoS). Efficient and reliable communication in wireless networks heavily relies on robust error correction mechanisms to ensure data integrity amidst varying channel conditions \cite{3GPPTR38212,lin2004error, chen2022standardization}. A key aspect of these mechanisms lies in the ability of decoders to provide soft output, furnishing a measure of confidence regarding the accuracy of decoded information \cite{lin2004error, chen2022standardization}. Traditionally, this measure has been facilitated by appending a CRC to transmitted messages, allowing for post-decoding verification \cite{hashimoto1997performance}, \cite{sauter2023error}. Consequently, the other widely common conjecture arises that with the 5G and 5G-beyond advent of 3GPP standards, characterized by ultra-reliable low-latency communication (URLLC) and the prevalence of short packets, the incorporation of a CRC can significantly impact the code's rate, prompting the exploration of alternative techniques for assessing decoding confidence.

%%
%% New Paragraph
%%
In addition to these two technological evidences mentioned above, GRAND (Guessing Random Additive Noise Decoding) offers an innovative solution for error correction in noisy communication channels \cite{duffy2019capacity,an2022keep}. As succinctly described in \secref{Sec:SystemModel:GRANDTransmission}, sequentially searching all possible erroneous bit patterns ordered based on their likelihood probabilities, GRAND intelligently guesses the error bit pattern affecting the transmitted binary bits\cite{duffy2021guessing,duffy2022ordered}. There exist GRAND algorithms achieving maximum-likelihood decoding for both hard and soft detection  \cite{solomon2020soft,duffy2021guessing,duffy2022ordered,duffy2023using}. As such, these algorithms demonstrate flexibility and energy efficiency in hardware implementations, enhancing  reliability and efficiency without adding significant complexity such that it makes them promising candidates for practical deployment\cite{riaz2021multi,riaz2023sub} by the development of highly accurate soft-output measures for GRAND \cite{condo2021high,abbas2020high,abbas2022high,condo2022fixed}. 

%%
%% New Paragraph
%%
To the best of our knowledge, the potential benefits of combining multihop, CRC, and GRAND technologies within 3GPP networks have not been fully explored in the literature, despite their individual importance for the future. This integrated approach has the potential to significantly improve both the speed and quality of transmissions. Our contributions are summarized as follows:
\begin{itemize}
\IEEEcompact
% %---------------------------------------------------
% % Item
% %---------------------------------------------------
\item[\ITMsymbol] We provide an overview and theoretical BER-cause analysis of GRAND-based multihop transmission, elucidating its underlying principles and operational mechanisms within GRAND-decoding algorithm.
%Overview and theoretical analysis of GRAND-based multihop transmission, clarifying its underlying principles and operational mechanisms.
% %---------------------------------------------------
% % Item
% %---------------------------------------------------
 \item[\ITMsymbol] We implement simulations for GRAND-based multihop transmission under two distinct scenarios: (1) where only the destination utilizes GRAND decoding, and (2) where both relays and the destination leverage it. We employ both Additive White Gaussian Noise (AWGN) and Rayleigh fading channel models. Through this simulation-based performance analysis, we meticulously evaluate the efficacy and viability of GRAND-based multihop transmission.
  % Implementation of simulations for GRAND-based multihop transmission under Additive White Gaussian Noise (AWGN) and Rayleigh fading channel models across two different scenarios. 
 % %---------------------------------------------------
% % Item
% %---------------------------------------------------
 \item[\ITMsymbol] Leveraging the simulation-based results for BER performance, we conduct a meticulous performance analysis to evaluate the efficacy and viability of GRAND-based multihop transmission. This analysis culminates in a robust conclusion that highlights the promising potential of this novel approach for enhancing communication speed and reliability in future wireless networks.
\end{itemize}

%%
%% New Paragraph
%%
The remainder of this paper is organized as follows: \secref{Sec:SystemModel} introduces the system model employed for GRAND-based single-hop and multihop transmission, followed by a theoretical BER-cause analysis of GRAND decoding in multihop scenarios. \secref{Sec:PerformanceAnalysisAndSimulationResults} delves into the performance analysis of GRAND for multihop transmission, presenting detailed simulation results. Finally, the paper concludes in \secref{Sec:Conclusion} by summarizing the key findings and their implications.

%%%%%%%%%%%%%%%%%%%%%%%%%%%%%%%%%%%%%%%%%%%%%%%%%%%%%%%%%%%%%%%%%%%%%%%%%%%%%%%%%%%%%%%%%%%%%%%%%%%%%%%%%%%%%%%%%%%%%%%%%%%%
%%   _________              __  .__               
%%  /   _____/ ____   _____/  |_|__| ____   ____  
%%  \_____  \_/ __ \_/ ___\   __\  |/  _ \ /    \ 
%%  /        \  ___/\  \___|  | |  (  <_> )   |  \
%% /_______  /\___  >\___  >__| |__|\____/|___|  /
%%         \/     \/     \/                    \/ 
\section{System Model} \label{Sec:SystemModel}
\IEEEcompact

%%   _________    ___.                         __  .__               
%%  /   _____/__ _\_ |__   ______ ____   _____/  |_|__| ____   ____  
%%  \_____  \|  |  \ __ \ /  ___// __ \_/ ___\   __\  |/  _ \ /    \ 
%%  /        \  |  / \_\ \\___ \\  ___/\  \___|  | |  (  <_> )   |  \
%% /_______  /____/|___  /____  >\___  >\___  >__| |__|\____/|___|  /
%%         \/          \/     \/     \/     \/                    \/ 
\subsection{Grand-based (Single-Hop) Transmission}\label{Sec:SystemModel:GRANDTransmission}
\IEEEcompact
Let us consider a single-input single-output (SISO) wireless communication, considered as a single-hop transmitting from transmitter to the receiver, over generalized fading channels. Then, the transmitted and received symbol blocks are related by 
\begin{equation}\label{Eq:TransmissionOverFadingChannels}
\mathbf{y}=\mathbf{H}\mathbf{x}+\mathbf{n},
\end{equation}
where the vector $\mathbf{x}\!=\![x_{1},x_{2},\cdots{x}_{m},\cdots,{x}_{M}]^{T}\!\in\!\mathbb{C}^{M}$ represents the transmitted block comprising $M$ modulated symbols. Similarly, $\mathbf{y}\!=\![y_{1},y_{2},\cdots,{y}_{M}]^{T}\!\in\!\mathbb{C}^{M}$ represents the receiving block comprising noisy $M$ modulated symbols. The transmitted block $\mathbf{x}$ is exposed to the channel fading characterized by diagonal matrix $\mathbf{H}\!=\![h_{ii}\!\neq\!0\,|\,h_{ij}\!=\!0, i\!\neq\!j]\!\in\!\mathbb{C}^{M\times{M}}$, indicating point-to-point fading. The vector $\mathbf{n}\!=\![n_{1},n_{2},\cdots,{n}_{M}]\!\in\!\mathbb{C}^{M}$ is the complex additive white Gaussian noise (AWGN). We assume, for the transmitted symbol block $\mathbf{x}$, the symbols $x_{m},(1\!\leq\!{m}\!\leq\!{M})$ are equiprobably taken from a normalized complex constellation $\mathcal{X}$ such that $\forall{m},x_{m}\in \mathcal{X}$ and $\mathbb{E}\left[\mathbf{x}^{*}\mathbf{x}\right]=1$. Each symbol is assumed carrying $q\!=\!\left\lceil\log_{2}\left(\left|\mathcal{X}\right|\right)\right\rceil$ bits. We can express the bit-representation of symbol $x_{m}$ as  
\begin{equation}
    \mathbf{c}_{m}=\left[{c}_{m1}, {c}_{m2}, \cdots ,{c}_{mq}\right]^{T}\in\mathbb{F}_{2}^{q}.
\end{equation}
Hence, each transmitted block carries $N=qM$ bits, denoted by 
\begin{equation}
    \mathbf{c}=\left[\mathbf{c}_{1},\mathbf{c}_{2}, \cdots, \mathbf{c}_{m}, \cdots \mathbf{c}_{M}\right]^{T}\in\mathbb{F}_{2}^{q\times{M}}.
\end{equation}
such that $|\mathbf{c}|=N$. Assuming no loss of generality, let us denote $\mathbf{c}$ as safeguarded by a CRC. This CRC can be uniquely denoted by an injective function $\mathrm{CRC}\!:\!\mathbb{F}_{2}^{K}\!\rightarrow\!\mathbb{F}_{2}^{M}$ with a code-rate of $R\!=\!K/N$ and it can be defined by a code-book, that is $\mathcal{C}\!\triangleq\!\{\mathbf{c}\left|\,\mathbf{c}\!=\!\mathrm{CRC}(\mathbf{b}),\mathbf{b}\!\in\!\mathbb{F}_{2}^{K}\right\}$,
which includes all possible code-words, where $\mathbf{b}$ denotes the uncoded bit vector. From the conventional perspective, conditioning on the received bits $\hat{\mathbf{c}}$, a maximum-likelihood (ML) decoder calculates the likelihood of each code-word, and chooses the code-word $\mathbf{c}^{\mathrm{ML}}$ having the highest likelihood, given $2^K$ code-words in set $\mathcal{C}$, that is
\begin{equation}\label{Eq:MLDecoder}
    \mathbf{c}^{\mathrm{ML}}=\argmax_{\mathbf{c}\in \mathcal{C}}
        \Pr(\text{received}~\hat{\mathbf{c}}\mid\text{transmitted}~\mathbf{c})
\end{equation}
where the received bits, denoted as $\hat{\mathbf{c}}$, may contain errors represented by $\mathbf{e}$, resulting from various sources such as additive noise and interference,  channel fading and hardware imperfections. These factors can distort the transmission, causing inconsistencies between the bit representations of transmitted block $\mathbf{x}\!\in\!\mathbb{C}^{M}$ and received block $\mathbf{y}\!\in\!\mathbb{C}^{M}$. Thus, we can express the received bits $\hat{\mathbf{c}}\!\in\!\mathbb{F}_{2}^{N}$ in relation to the transmitted bits $\mathbf{c}$ and the erroneous bits $\mathbf{e}\!\in\!\mathbb{F}_{2}^{N}$, also referred to as additive binary noise, that is $\hat{\mathbf{c}}=\mathbf{c}\oplus\mathbf{e}$, where $\oplus$ denotes the modulo-$2$ XOR operation. Upon using $\mathbf{c}\oplus\mathbf{c}\!=\!\mathbf{0}$, we can express the additive binary noise $\mathbf{e}$ in terms of the transmitted and received bits as follows $\mathbf{e}=\hat{\mathbf{c}}\oplus\mathbf{c}$ and simplify the ML decoder, given by \eqref{Eq:MLDecoder}, to 
\begin{subequations}\label{Eq:MLDecoderSimplification}
\setlength\arraycolsep{1.4pt}
\begin{eqnarray}
\label{Eq:MLDecoderSimplificationA}
\mathbf{c}^{\mathrm{ML}}
    &\triangleq&\argmax_{\mathbf{c}\in \mathcal{C}}
        \Pr(\text{received}~\hat{\mathbf{c}}\mid\text{transmitted}~\mathbf{c}),\\
\label{Eq:MLDecoderSimplificationB}
    &=&\argmax_{\mathbf{c}\in \mathcal{C}}
        \Pr(\hat{\mathbf{c}}\oplus\mathbf{c}\mid\text{transmitted}~\mathbf{c}),\\
\label{Eq:MLDecoderSimplificationC}
    &\overset{(a)}{=}&\argmax_{\mathbf{c}\in \mathcal{C}}
        \frac{\Pr(\hat{\mathbf{c}}\oplus\mathbf{c},\text{transmitted}~\mathbf{c})}{\Pr(\text{transmitted}~\mathbf{c})},\\
\label{Eq:MLDecoderSimplificationD}
    &\overset{(b)}{=}&\argmax_{\mathbf{c}\in \mathcal{C}}
        \Pr(\hat{\mathbf{c}}\oplus\mathbf{c},\text{transmitted}~\mathbf{c}),\\
\label{Eq:MLDecoderSimplificationE}
    &=&\argmax_{\mathbf{c}\in \mathcal{C}}
        \Pr(\mathbf{e}=\hat{\mathbf{c}}\oplus\mathbf{c}),   
\end{eqnarray}
\end{subequations}
where, in the step from $(a)$ to $(b)$, we assign $\Pr(\mathbf{c}_m)\!=\!1/|\mathcal{C}|$ for all ${c}_{m}\!\in\!\mathcal{C}$, thus ignoring the denominator $\Pr(\text{transmitted}~\mathbf{c})$ due to the assumption of equiprobable modulation symbols in transmission. Instead of searching through $\mathcal{C}$ for code-words in ML decoding, \eqref{Eq:MLDecoderSimplificationE} proposes, exploiting the error-detection capability of CRC, a search for the best possible additive binary noise $\mathbf{e}\!=\![e_1,e_2,\ldots,e_N]^{T}\!\in\!\mathbb{F}_{2}^{N}$ whose random nature corrupts the transmitted bits $\mathbf{c}$  such that all possible error patterns are ordered with non-increasing probability in the set $\mathcal{E}$. This insightful approach, known as GRAND \cite{duffy2019capacity}, actively defines the GRAND decoder as
\begin{equation}\label{Eq:GRANDDecoder}
    \mathbf{e}^{\mathrm{GRAND}}=
        \argmax_{\mathbf{e}\in \mathcal{E}}
            \Pr(\mathrm{CRC}(\mathbf{e}\oplus\hat{\mathbf{c}})~\text{succeeds})
\end{equation}
which actively seeks to estimate the optimal binary additive noise, maximizing the probability of successful CRC decoding, thus enabling the recovery of the transmitted bits with maximum accuracy, that is  
\begin{equation}
    \mathbf{c}^{\mathrm{GRAND}}=\mathbf{e}^{\mathrm{GRAND}}\oplus\hat{\mathbf{c}}. 
\end{equation}
which is the transmitted bits detected by GRAND decoder. 

%%
%% New Paragraph
%%
It is worth mentioning that, since the entropy of binary additive noise has generally lower entropy than information bits $\mathbf{c}$, employing GRAND reduces decoding complexity while maintains optimal error exponents. This efficiency boost relies on two main strategies, one of which is sorting the binary additive noises (error patterns) in set $\mathcal{E}$ by decreasing probability, and the other one is stopping error-pattern guessing after a predetermined computational limit \cite{duffy2019capacity}. In this paper, we consider GRAND decoding and \emph{ordered-reliability-bits}-based GRAND (ORBGRAND), that is 
\begin{itemize}
% %---------------------------------------------------
\IEEEcompact
%---------------------------------------------------
% Item
%---------------------------------------------------
\item[\ITMsymbol] $\mathrm{GRAND~\text{(hard~detection)}}$. The classic hard-detection decoder \cite{duffy2019capacity} prioritizing additive binary noises  (also referred to as error patterns) by their weights, and in querying, favoring those with lower weights. If error patterns have equal weight, then their order is arbitrary among themselves.
%---------------------------------------------------
% Item
%---------------------------------------------------
\item[\ITMsymbol] $\mathrm{ORBGRAND~\text{(soft~detection)}}$. The soft-detection decoder \cite{duffy2021ordered} utilizing bit reliability metrics extracted from the statistical soft-information of the received bits, then calculating the probabilities of possible error patterns and prioritizing the least reliable ones for correction. Through an iterative process of flipping bits and checking for valid code-words, it corrects errors while remaining efficient for hardware implementation.
\end{itemize}

%%   _________    ___.                         __  .__               
%%  /   _____/__ _\_ |__   ______ ____   _____/  |_|__| ____   ____  
%%  \_____  \|  |  \ __ \ /  ___// __ \_/ ___\   __\  |/  _ \ /    \ 
%%  /        \  |  / \_\ \\___ \\  ___/\  \___|  | |  (  <_> )   |  \
%% /_______  /____/|___  /____  >\___  >\___  >__| |__|\____/|___|  /
%%         \/          \/     \/     \/     \/                    \/ 
\subsection{Grand-based Multihop Transmission}\label{Sec:SystemModel:GRANDMultihopTransmission}\IEEEcompact
Let us consider the system model of multihop ($L$-hop) decode-and-forward (DF) transmission, in which the information transmission from the source node $S$ to the destination node $D$ involves $L\!-\!1$ relay nodes, denoted as $R_1, R_2,\ldots,R_{L-1}$. This scheme employs decode-and-forward methodology, where each relay node decodes the received signal before re-encoding and forwarding it to the next hop. The sequential transmissions over generalizing fading environments among sequential nodes are assumed experiencing channel impairments such as path loss, shadowing, and multipath fading, and also assumed subjected to additive white Gaussian noise, which affect the reliability and quality of the transmitted information. 

%%
%% New Paragraph
%%
Exploiting GRAND to mitigate the error patterns in the received bits, we propose two approaches to enhance the robustness of multihop transmission over generalized fading channels:

%%  _________                                .__          ____ 
%% /   _____/ ____  ____   ____ _____ _______|__| ____   /_   |
%% \_____  \_/ ___\/ __ \ /    \\__  \\_  __ \  |/  _ \   |   |
%% /        \  \__\  ___/|   |  \/ __ \|  | \/  (  <_> )  |   |
%%/_______  /\___  >___  >___|  (____  /__|  |__|\____/   |___|
%%        \/     \/    \/     \/     \/                        
\noindent
\emph{(Scenario \#1).} In the first scenario, GRAND technology is only employed at the destination node $D$ to estimate and mitigate additive binary noise bits within the bits received from the source node $S$ through the transmission chain of relay nodes. Let $\mathbf{c}$ represent a block of code-words. The source node $S$ modulates $\mathbf{c}$, safeguarded by a CRC code, and then obtains the symbol block $\mathbf{x}$, sending it to the destination node $D$ through a series of relay nodes. Using the notation from \eqref{Eq:TransmissionOverFadingChannels}, the relay node $R_{\ell}$ receives the noisy symbol block $\mathbf{y}_{\ell}$ from the previous relay node $R_{\ell-1}$, where $R_{0}$ stands for the source node $S$. The noisy symbol block $\mathbf{y}_{\ell}$ is written as
\begin{equation}
\mathbf{y}_{\ell}=\mathbf{H}_{\ell}\mathbf{x}_{\ell-1}+\mathbf{n}_{\ell},     
\end{equation}
where $\mathbf{x}_{\ell}$ denotes the symbol block transmitted by relay node $R_{\ell}$ and $\mathbf{x}_{0}$ denotes the symbol block $\mathbf{x}$ transmitted by source node $S$. The relay node $R_{\ell}$ decodes (demodulates) $\mathbf{y}_{\ell}$ as the bit representation $\hat{\mathbf{c}}_{\ell}$, described as 
\begin{equation}
\hat{\mathbf{c}}_{\ell}=\hat{\mathbf{c}}_{\ell-1}\oplus\mathbf{e}_{\ell},
\end{equation}
where $\mathbf{e}_{\ell}$ represents the binary noise (error pattern) during transmission from relay node $R_{\ell-1}$ to relay node $R_{\ell}$, and hence $\hat{\mathbf{c}}_{0}$ represents the bit representation of the symbol block $\mathbf{c}$ transmitted by the source node $S$. At the end of all transmissions through the relay node chain, the bit representation $\hat{\mathbf{c}}$ received by the destination node $D$ is readily obtained as 
\begin{equation}
\hat{\mathbf{c}}=\mathbf{c}\oplus\mathbf{e}_{\scriptscriptstyle\sum},
\end{equation}
where $\mathbf{e}_{\scriptscriptstyle\sum}=\mathbf{e}_{1}\oplus\mathbf{e}_{2}\oplus\cdots\oplus\mathbf{e}_{L}$ denotes the overall binary additive noise as a result of decode-and-forward multihop transmission. Finally, GRAND technology is employed by the destination node $D$ to mitigate the additive binary noise bits $\mathbf{e}_{\scriptscriptstyle\sum}$ as follows
\begin{equation}
    \mathbf{e}^{\mathrm{GRAND}}=
        \argmax_{\mathbf{e}\in \mathcal{E}}
            \Pr(\mathrm{CRC}(\mathbf{e}\oplus\hat{\mathbf{c}})~\text{succeeds}),
\end{equation}
which actively seeks to estimate the optimal binary additive noise, maximizing the probability $\Pr(\mathbf{e}_{\scriptscriptstyle\sum}\!=\!\mathbf{e}^{\mathrm{GRAND}})$, thus enabling the recovery of the transmitted bits with maximum accuracy, that is $\mathbf{c}^{\mathrm{GRAND}}\!=\!\mathbf{e}^{\mathrm{GRAND}}\oplus\hat{\mathbf{c}}$, which are the transmitted bits detected by the GRAND decoder. Accordingly, the BER is obtained by
\begin{subequations}
\setlength\arraycolsep{1.4pt}
\begin{eqnarray}
\!\!\!\!\!\!P_{\mathrm{BER}}
&=&\mathbb{E}\Bigl[\mathbf{e}_{\scriptscriptstyle\sum}\!\oplus\!\mathbf{e}^{\mathrm{GRAND}}\Bigr],\\
&=&\mathbb{E}\Bigl[\mathbf{e}_{\scriptscriptstyle\sum}\!\oplus\!
        \argmax_{\mathbf{e}\in \mathcal{E}}
                \Pr(\mathrm{CRC}(\mathbf{e}\!\oplus\!\hat{\mathbf{c}})~\text{succeeds})\Bigr].~~~
\end{eqnarray}
\end{subequations}

%%   _________                                .__         ________  
%%  /   _____/ ____  ____   ____ _____ _______|__| ____   \_____  \ 
%%  \_____  \_/ ___\/ __ \ /    \\__  \\_  __ \  |/  _ \   /  ____/ 
%%  /        \  \__\  ___/|   |  \/ __ \|  | \/  (  <_> ) /       \ 
%% /_______  /\___  >___  >___|  (____  /__|  |__|\____/  \_______ \
%%         \/     \/    \/     \/     \/                          \/
\noindent
\emph{(Scenario \#2).} In the second scenario, each relay node $R_{\ell}$ applies GRAND decoding to the bits received from the preceding relay node $R_{\ell-1}$, where relay $R_{0}$ is the source node $S$. This distributed noise mitigation approach extends the noise estimation and discrimination capabilities throughout the transmission chain. At each relay node, GRAND technology is utilized to accurately estimate and suppress additive binary noise bits from the received signal before forwarding it to the next hop. This iterative noise mitigation process at each relay node contributes to improved signal quality and robustness against channel impairments. Another limitation of CRC is its vulnerability to undetected errors. Despite being designed to detect errors, CRC may fail to identify certain types of errors. For example, if the number of corrupted bits in a message equals the size of the CRC polynomial, the errors will cancel each other out, resulting in a false positive. This means that CRC is not always able to accurately identify when errors have occurred, potentially leading to data corruption or transmission issues. In accordance with the stages, except each relay node applies GRAND decoding, at the end of all transmissions through the relay node chain, the bit block $\hat{\mathbf{c}}$ received by destination node $D$ is obtained as $\hat{\mathbf{c}}\!=\!\mathbf{c}\oplus\mathbf{e}_{\scriptscriptstyle\sum}$, where $\mathbf{e}_{\scriptscriptstyle\sum}$ is given~by 
\begin{equation}
\mathbf{e}_{\scriptscriptstyle\sum}=
    \widehat{\mathbf{e}}^{\mathrm{GRAND}}_{1}\oplus
        \widehat{\mathbf{e}}^{\mathrm{GRAND}}_{2}\oplus
            \cdots\oplus
                \widehat{\mathbf{e}}^{\mathrm{GRAND}}_{L-1}
                \oplus
                \mathbf{e}_{L}.
\end{equation}
where $\widehat{\mathbf{e}}^{\mathrm{GRAND}}_n$, $(1\leq\ell\leq{L})$ denote undetectable binary noises that pass the CRC check, meaning they are not flagged as errors, and $\mathbf{e}_{L}$ is the binary noise of the last hop in the transmission. The destination node detects the additive binary noise $\mathbf{e}_{\scriptscriptstyle\sum}$ as 
\begin{equation}
    \mathbf{e}^{\mathrm{GRAND}}_{L}=
        \argmax_{\mathbf{e}\in \mathcal{E}}
            \Pr(\mathrm{CRC}(\mathbf{e}\oplus\hat{\mathbf{c}})~\text{succeeds}),
\end{equation}
and then the transmitted bit block as $\mathbf{c}^{\mathrm{GRAND}}\!=\!\allowbreak\hat{\mathbf{c}}\oplus\mathbf{e}^{\mathrm{GRAND}}_{L}$. Accordingly, the BER is calculated as 
\begin{subequations}
\label{Eq:BERExpressionForScenario2}
\setlength\arraycolsep{1.4pt}
\begin{eqnarray}
\label{Eq:BERExpressionForScenario2A}
\!\!\!\!\!\!P_{\mathrm{BER}}
&=&\mathbb{E}\Bigl[\mathbf{c}\oplus\mathbf{c}^{\mathrm{GRAND}}\Bigr],\\
\label{Eq:BERExpressionForScenario2B}
&=&\mathbb{E}\Bigl[\mathbf{e}_{\scriptscriptstyle\sum}\!\oplus\!
        \argmax_{\mathbf{e}\in \mathcal{E}}
                \Pr(\mathrm{CRC}(\mathbf{e}\!\oplus\!\hat{\mathbf{c}})~\text{succeeds})\Bigr],\\ 
\label{Eq:BERExpressionForScenario2C}
&=&\mathbb{E}\Bigl[\widehat{\mathbf{e}}^{\mathrm{GRAND}}_{1}\oplus\widehat{\mathbf{e}}^{\mathrm{GRAND}}_{2}\oplus\cdots\oplus\widehat{\mathbf{e}}^{\mathrm{GRAND}}_{L}\Bigr].~~~
\end{eqnarray}
\end{subequations}
where $\widehat{\mathbf{e}}^{\mathrm{GRAND}}_{L}$ is the undetectable noise at the last CRC check.

%%
%% New Paragraph
%%
The two scenarios mentioned above involve transmission over generalized fading channels. The scenario choice between scenarios depends on trade-offs between computational complexity, latency, and adaptability to channel variations. Scenario \#1 simplifies transmission but places more noise-mitigation responsibility on the destination node. In contrast, Scenario \#2 shares noise mitigation among all relay nodes, bolstering system resilience while potentially elevating computational overhead.

%% ___________.__                             
%% \_   _____/|__| ____  __ _________   ____  
%%  |    __)  |  |/ ___\|  |  \_  __ \_/ __ \ 
%%  |     \   |  / /_/  >  |  /|  | \/\  ___/ 
%%  \___  /   |__\___  /|____/ |__|    \___  >
%%      \/      /_____/                    \/ 
%% GRAND AWGN
\begin{figure}[t]
  \centering
  \begin{minipage}{0.4\textwidth}
  \centering
    \includegraphics[width=\textwidth,trim=0.5cm 0.0cm 1cm 0.5cm, clip, keepaspectratio]{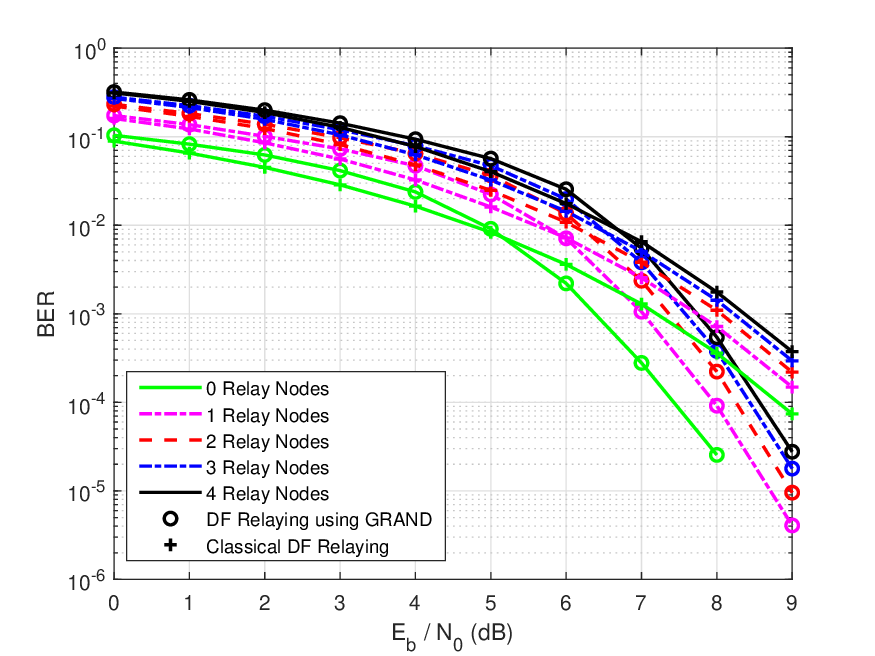}
    \subcaption{Scenario \#1.}
    \label{GRANDInAWGNChannelsScenario1}
  \end{minipage}
  \\[1mm]%\hfill
  \begin{minipage}{0.4\textwidth}
  \centering
    \includegraphics[width=\textwidth,trim=0.5cm 0.0cm 1cm 0.5cm, clip, keepaspectratio]{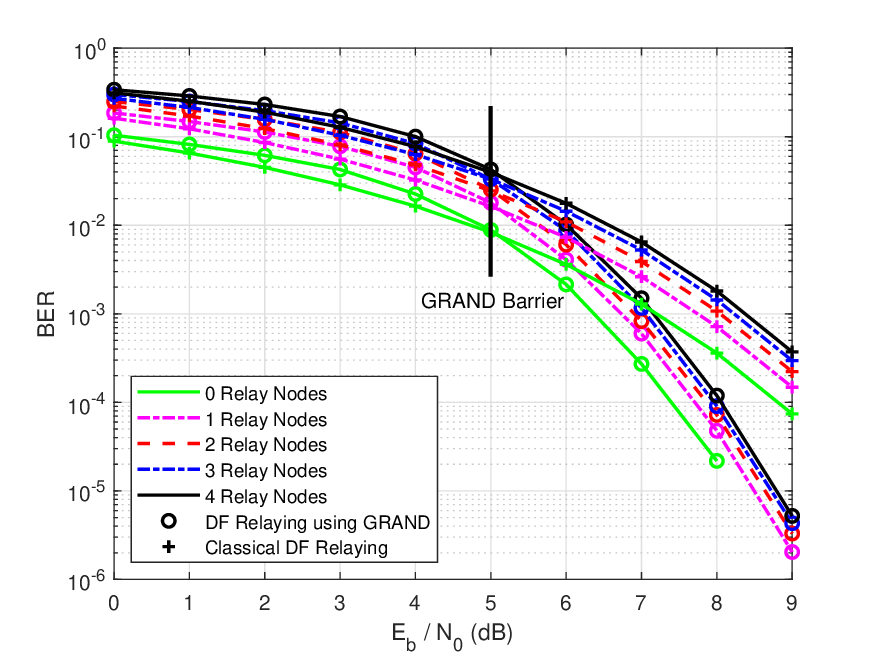}
    \subcaption{Scenario \#2.}
    \label{GRANDInAWGNChannelsScenario2}
  \end{minipage}
  \\[0mm]
  \setlength{\belowcaptionskip}{-4mm}
  \caption{\small BER performance of BPSK modulation with GRAND decoding in decode-and-forward multihop transmission scheme over AWGN channels}
  \label{Fig:grandawgn}
  %\vspace{-5mm}
\end{figure}

%% ___________.__                             
%% \_   _____/|__| ____  __ _________   ____  
%%  |    __)  |  |/ ___\|  |  \_  __ \_/ __ \ 
%%  |     \   |  / /_/  >  |  /|  | \/\  ___/ 
%%  \___  /   |__\___  /|____/ |__|    \___  >
%%      \/      /_____/                    \/ 
%% GRAND Rayleigh
\begin{figure}[t]
  \centering
  \begin{minipage}{0.4\textwidth}
    \includegraphics[width=\textwidth,trim=0.5cm 0.0cm 1cm 0.5cm, clip, keepaspectratio]{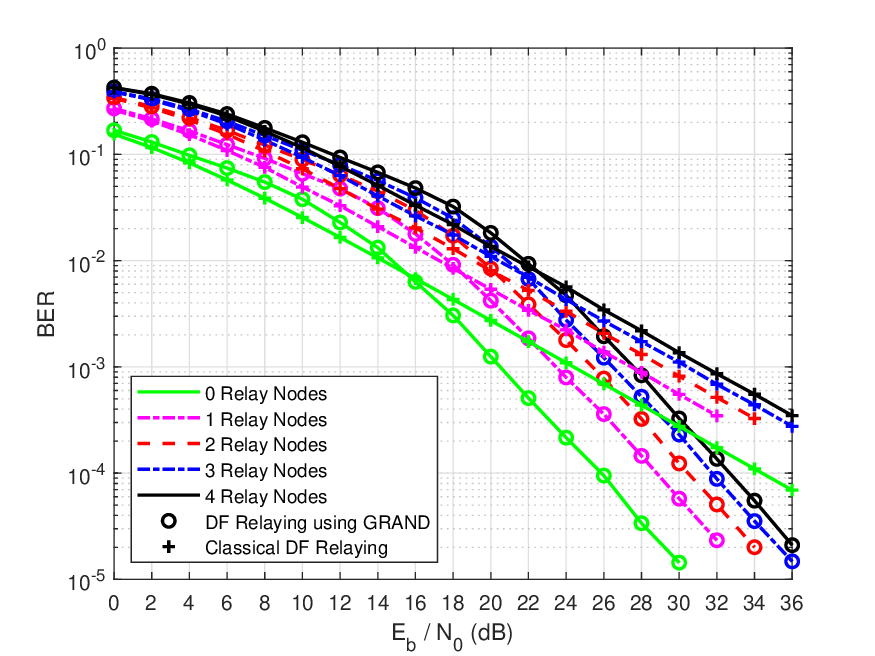}
    \subcaption{Scenario \#1.}
    \label{GRANDInRayleighFadingChannelsScenario1}
  \end{minipage}
  \\[1mm]%\hfill
  \begin{minipage}{0.4\textwidth}
    \includegraphics[width=\textwidth,trim=0.5cm 0.0cm 1cm 0.5cm, clip, keepaspectratio]{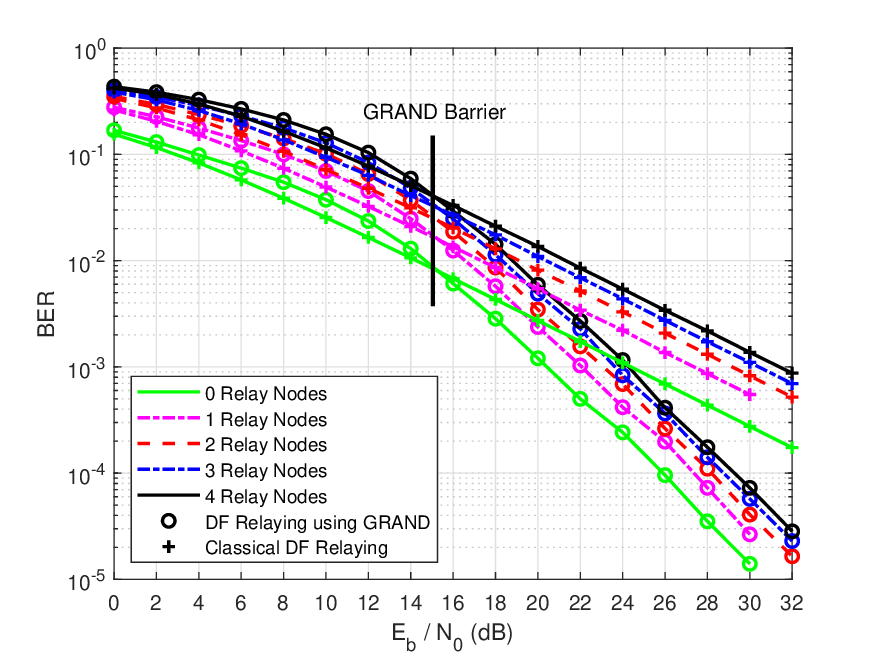}
    \subcaption{Scenario \#2.}
    \label{Fig:GRANDInRayleighFadingChannelsScenario2}
  \end{minipage}
  \\[0mm]
  \setlength{\belowcaptionskip}{-5mm}
  \caption{\small BER performance of BPSK modulation with GRAND decoding in decode-and-forward multihop transmission scheme over Rayleigh fading channels}
  \label{Fig:grandrayleigh}
  %\vspace{-5mm}
\end{figure}

%% ___________.__                             
%% \_   _____/|__| ____  __ _________   ____  
%%  |    __)  |  |/ ___\|  |  \_  __ \_/ __ \ 
%%  |     \   |  / /_/  >  |  /|  | \/\  ___/ 
%%  \___  /   |__\___  /|____/ |__|    \___  >
%%      \/      /_____/                    \/ 
%% ORBGRAND AWGN 
\begin{figure}[t]
  \centering
  \begin{minipage}{0.4\textwidth}
    \includegraphics[width=\textwidth,trim=0.5cm 0.0cm 1cm 0.5cm, clip, keepaspectratio]{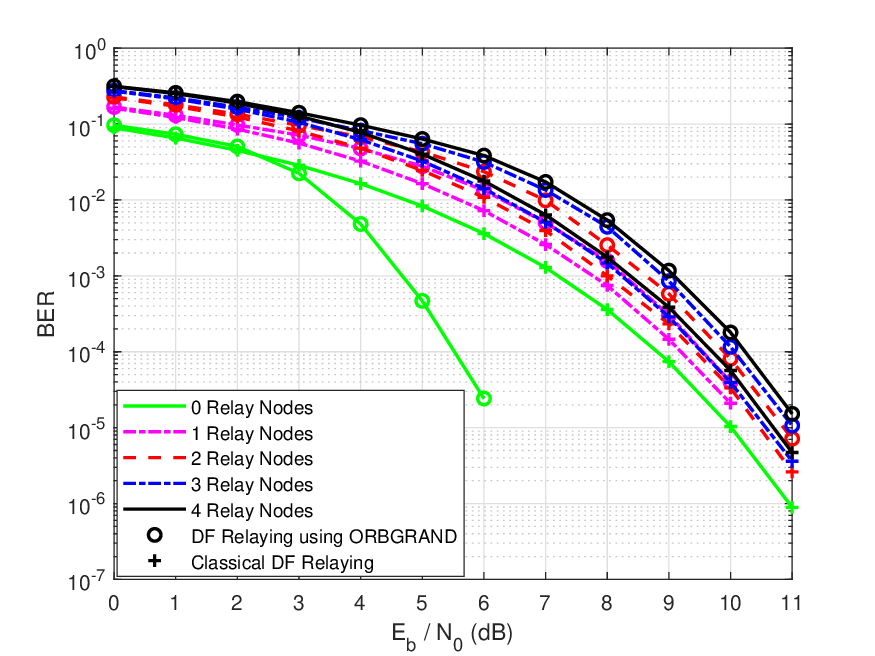}
    \subcaption{Scenario \#1.}
    \label{ORBGRANDInAWGNChannelssc1}
  \end{minipage}
  \\[1mm]%\hfill
  \begin{minipage}{0.4\textwidth}
    \includegraphics[width=\textwidth,trim=0.5cm 0.0cm 1cm 0.5cm, clip, keepaspectratio]{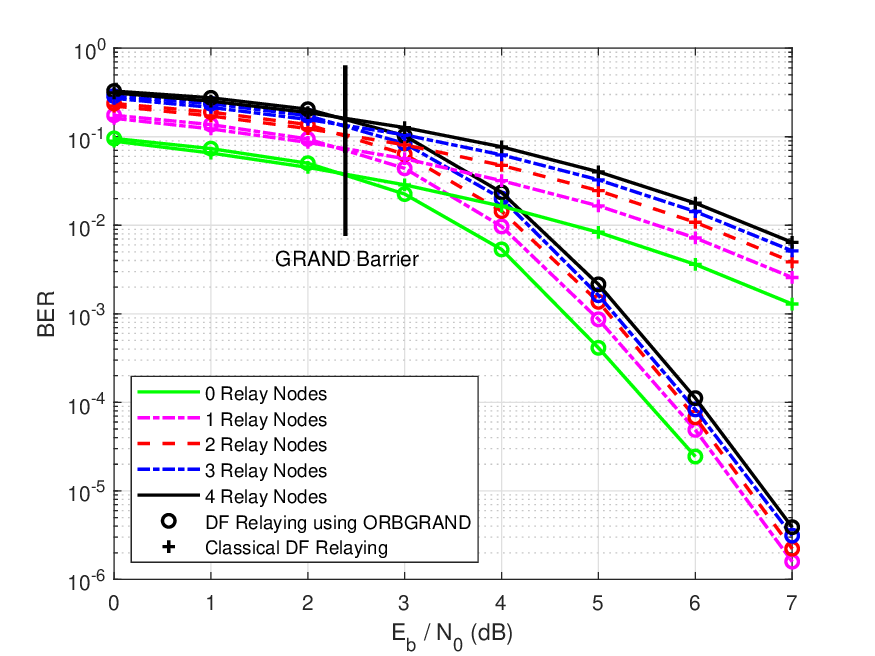}
    \subcaption{Scenario \#2.}
    \label{ORBGRANDInAWGNChannelsScenario2}
  \end{minipage}
  \\[0mm]
  \setlength{\belowcaptionskip}{-5mm}
  \caption{\small BER performance of BPSK modulation with ORBGRAND decoding in decode-and-forward multihop transmission scheme over AWGN channels}
  \label{Fig:ORBGRANDInAWGNChannels}
  %\vspace{-5mm}
\end{figure}

%% ___________.__                             
%% \_   _____/|__| ____  __ _________   ____  
%%  |    __)  |  |/ ___\|  |  \_  __ \_/ __ \ 
%%  |     \   |  / /_/  >  |  /|  | \/\  ___/ 
%%  \___  /   |__\___  /|____/ |__|    \___  >
%%      \/      /_____/                    \/ 
%% ORBGRAND Rayleigh
\begin{figure}[t]
  \centering
  \begin{minipage}{0.4\textwidth}
    \includegraphics[width=\textwidth,trim=0.5cm 0.0cm 1cm 0.5cm, clip, keepaspectratio]{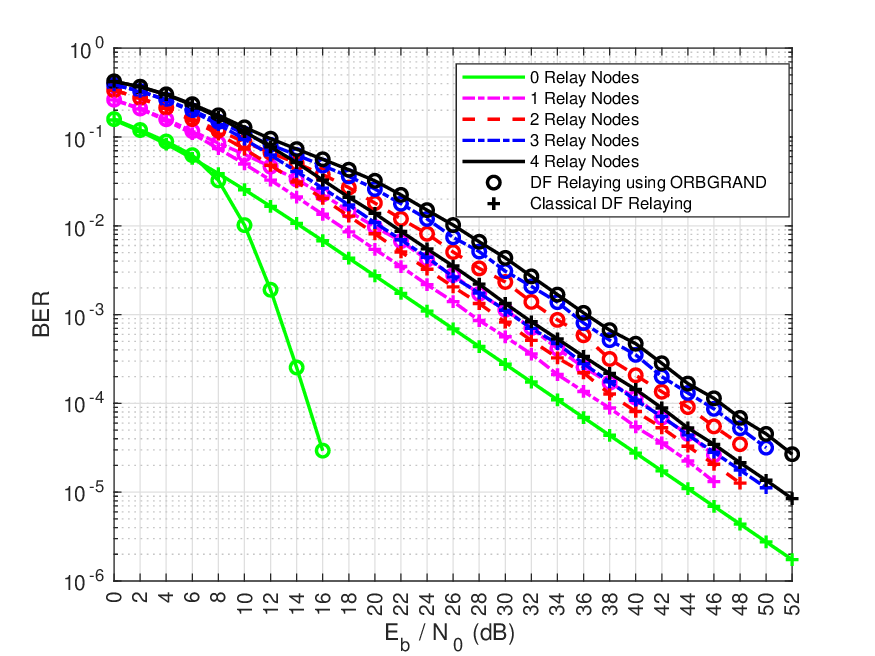}
    \subcaption{Scenario \#1.}
    \label{Fig:ORBGRANDInRayleighChannelssc1}
  \end{minipage}
  \\[2mm]%\hfill
  \begin{minipage}{0.4\textwidth}
    \includegraphics[width=\textwidth,trim=0.5cm 0.0cm 1cm 0.5cm, clip, keepaspectratio]{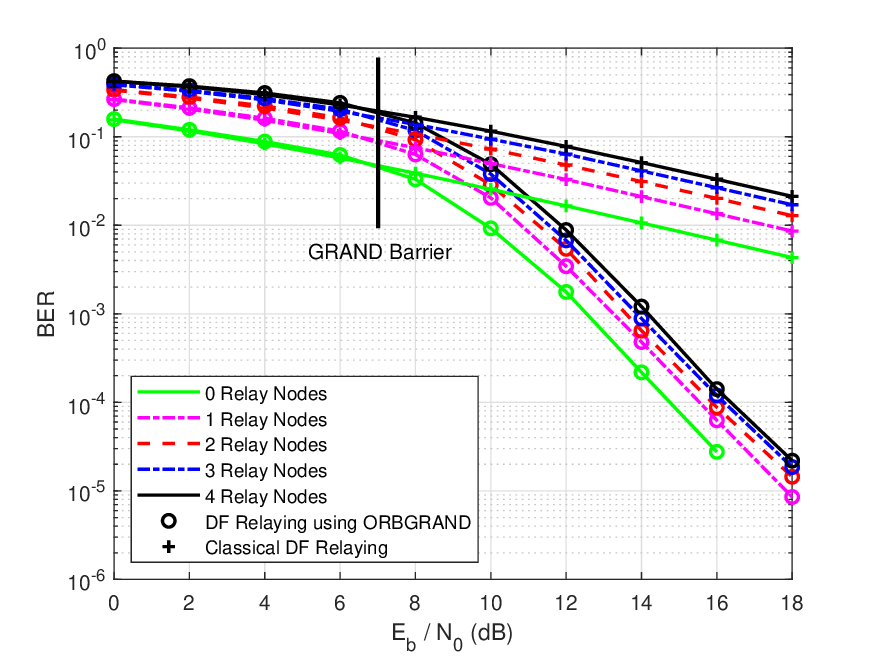}
    \subcaption{Scenario \#2.}
    \label{ORBGRANDInRayleighFadingChannelsScenario2}
  \end{minipage}
  \\[1mm]
  \setlength{\belowcaptionskip}{-5mm}
  \caption{\small BER performance of BPSK modulation with ORBGRAND decoding in decode-and-forward multihop transmission scheme over Rayleigh fading channels}
  \label{Fig:ORBGRANDInRayleighChannels}
  %\vspace{-5mm}
\end{figure}

%%   _________              __  .__               
%%  /   _____/ ____   _____/  |_|__| ____   ____  
%%  \_____  \_/ __ \_/ ___\   __\  |/  _ \ /    \ 
%%  /        \  ___/\  \___|  | |  (  <_> )   |  \
%% /_______  /\___  >\___  >__| |__|\____/|___|  /
%%         \/     \/     \/                    \/ 
\section{Performance Results}\label{Sec:PerformanceAnalysisAndSimulationResults}
In the following, we provide detailed simulation results by presenting BER curves with respect to $E_b/N_0$. We evaluate scenarios described in subsection \ref{Sec:SystemModel:GRANDMultihopTransmission} for various numbers of relay nodes ($L\!=\!0,1,2,3,4$). BPSK (Binary Phase-Shift Keying) modulation is used as the modulation scheme. We employ $\mathrm{CRC}\!-\!12$ with the Koopman polynomial $\mathrm{0\!\!\times\!\!8F3}$ to encode messages of length $k\!=\!116$ bits into $n\!=\!128$ bit code-words. Performance is evaluated under two channel conditions: an AWGN channel with no fading and a Rayleigh fading channel. In both cases, we assume perfect Channel State Information (CSI) is available at the relay nodes and the destination node. Specifically, in our calculations and simulations, we compute the variance $N_0$ for AWGN from 
\begin{equation}
\mathrm{Signalling~SNR}=({E_{b}}/{N_0})\!\times\!({k}/{n})\!\times\!{M}_0,
\end{equation}
where ${E_{b}}$ denotes the bit energy, and $M_{0}$ denotes modulation order (i.e., $M_{0}\!=\!1$ for BPSK).

%%
%% New Paragraph
%%
Scenario~\#1 investigates the GRAND decoding in a multihop transmission system, where relay nodes $R_1, R_2, ..., R_L$ receive symbol blocks from the preceding node, with $R_0$ representing the source. The relay nodes then demodulate the received signals, extract the bits while preserving the appended CRC bits for error detection.
Subsequently, the extracted bits are remodulated and forwarded to the next relay node, with $R_{L+1}$ representing the destination node. Notably, only the destination node utilizes the GRAND decoder, not the individual relays. \figref{GRANDInAWGNChannelsScenario1} and \figref{GRANDInRayleighFadingChannelsScenario1} depict the BER performance curves for BPSK modulation over AWGN channels and Rayleigh fading channels, respectively. As expected, the 
BER increases with the number of hops. Further, the figures clearly illustrate a significant improvement in BER performance when the GRAND decoder is employed at the destination. For comparative purposes in \figref{GRANDInAWGNChannelsScenario1}, consider a single-hop communication scenario (without relays) and the GRAND decoder at the destination. To achieve a BER of approximately $10^{-4}$ at an SNR of $8.3$ dB, the single-hop case requires the same performance as a $5$-hop transmission (using $4$ relays) with the GRAND decoder used at the destination. Similar deductions for the advantages of GRAND decoding at the destination also be inferred from \figref{GRANDInRayleighFadingChannelsScenario1} for Rayleigh fading channels.  

%%
%% New Paragraph
%%
In contrast to Scenario \#1, which employed GRAND decoding solely at the destination node, Scenario \#2 investigates the application of GRAND decoding also at all relay nodes within a multihop transmission, where each relay node and the destination node $D$ utilize GRAND decoding. Scenario \#2 exhibits intriguing phenomenon, evident in \figref{GRANDInAWGNChannelsScenario2} and \figref{Fig:GRANDInRayleighFadingChannelsScenario2}, which depict the BER performance curves for AWGN channels and Rayleigh fading channels, respectively. Interestingly, unlike the BER  curves  in Scenario \#1, the BER curves in Scenario \#2 intersect at a single SNR value, regardless of the number of relays, with those performance curves of classical decode-and-forward multihop transmission. This Single SNR value, which we termed the \emph{GRAND barrier}, is irrespective of the number of relays. This phenomenon occurs because the GRAND decoding at the destination node cannot detect specific error patterns that occur during decoding at each relay node. These undedectable error patterns, as expressed in \eqref{Eq:BERExpressionForScenario2C}, manifest as modulo-2 sums of error patterns that elude detection by the CRC employed at relay nodes in multihop transmission; hence, they emerge as a GRAND barrier in Scenario \#2. This barrier reveals the SNR value at which the GRAND decoding begins to outperform the case where it is not used. \figref{GRANDInAWGNChannelsScenario2} and \figref{Fig:GRANDInRayleighFadingChannelsScenario2} show that the GRAND barrier is approximately $E_b/N_0\!=\!5$ dB for AWGN channels and approximately $E_b/N_0\!=\!15$ dB for Rayleigh fading channels. Further, at SNR values exceeding the GRAND barrier, employing GRAND decoding at each relay node significantly enhances the BER performance. For instance, while the BER performance for a $5$-hop transmission with GRAND decoding at $8$ dB SNR is $2\!\times\!{10}^{-3}$, it becomes approximately $2\!\times\!{10}^{-4}$ when the GRAND decoding is employed at each relay. However, for the SNR below the GRAND barrier, employing GRAND decoding at each relay may deteriorate performance, thus, it is recommended for designing future multihop protocols within 3GPP standards to disable GRAND decoding at the relay nodes within this SNR range.

%%
%% New Paragraph
%%
In addition, the BER curves of multihop transmission using ORBGRAND decoding are depicted in \figref{Fig:ORBGRANDInAWGNChannels} for AWGN channels and \figref{Fig:ORBGRANDInRayleighChannels} for Rayleigh fading channels. Each figure shows, for Scenario \#1, ORBGRAND performs poorly, stemming from possible degradation in bit reliability calculated at the destination node, when intermediate relay nodes participate in transmission. Further, for Scenario \#2, the GRAND Barrier exists at around $2.4$ dB and $6.5$ dB for AWGN channels and Rayleigh fading channels, respectively. ORBGRAND outperforms GRAND, exhibiting lower values and a more significant improvement in BER performance.

%%   _________              __  .__               
%%  /   _____/ ____   _____/  |_|__| ____   ____  
%%  \_____  \_/ __ \_/ ___\   __\  |/  _ \ /    \ 
%%  /        \  ___/\  \___|  | |  (  <_> )   |  \
%% /_______  /\___  >\___  >__| |__|\____/|___|  /
%%         \/     \/     \/                    \/ 
\section{Conclusion} \label{Sec:Conclusion}
This paper investigates the performance of two GRAND variants, hard-detection GRAND and soft-detection ORBGRAND, in decode-and-forward multihop transmission system over generalized fading channels, with an emphasis on \emph{``GRAND barrier''} phenomenon. Simulation results show the effectiveness of these GRAND variants in different scenarios. Optimizing GRAND for relaying technlology can help explore new scenarios, especially for future 3GPP networks, by leveraging the GRAND barrier.

%%%%%%%%%%%%%%%%%%%%%%%%%%%%%%%%%%%%%%%%%%%%%%%%%%%%%%%%%%%%%%%%%%%%%%%%%%%%%%%%%%%%%%%%%%%%%%%%%%%%%%%%%%%%%%%%%%%%%%%%%%%%
%%__________._____.   .__  .__                                  .__
%%\______   \__\_ |__ |  | |__| ____   ________________  ______ |  |__ ___.__.
%% |    |  _/  || __ \|  | |  |/  _ \ / ___\_  __ \__  \ \____ \|  |  <   |  |
%% |    |   \  || \_\ \  |_|  (  <_> ) /_/  >  | \// __ \|  |_> >   Y  \___  |
%% |______  /__||___  /____/__|\____/\___  /|__|  (____  /   __/|___|  / ____|
%%        \/        \/              /_____/            \/|__|        \/\/
%%
%\bibliography{IEEEfull,bozkurt_yilmaz_grand_based_DF_multihop_transmission}
\bibliography{IEEEabrv,bozkurt_yilmaz_grand_based_DF_multihop_transmission}
\bibliographystyle{IEEEtran}
%--------------------------------------------------------------------------------------------------------------------------
\end{document}